\documentclass{aa}
\usepackage[varg]{txfonts}
\usepackage{graphicx}
\graphicspath{{./}{figures/}}
\usepackage{subfigure}
\bibpunct{(}{)}{;}{a}{}{,} 

\begin{document}

\title{Luminosity selection for gamma-ray bursts}
\author{S. Banerjee \inst{\ref{inst1}, \ref{inst2}}
            \and D. Guetta\inst{\ref{inst3}}}

\institute {
Inst. for Quantum Gravity, FAU Erlangen-Nuremberg,
Staudtstr. 7, 91058 Erlangen, Germany \label{inst1}
\and 
Department of Physics, Ben-Gurion University, P.O.Box 653, Beer-Sheva 84105 Israel \label{inst2}
\and 
Department of Physics, Ariel University, Ariel, Israel \label{inst3}}
\date{12 November 2021 / 18 March 2022}



\abstract
{{\bf Aim:} There exists inevitable scatter in the intrinsic luminosity of gamma-ray bursts (GRBs). If there is relativistic beaming in the source, viewing angle variation necessarily  introduces  variation in the intrinsic luminosity function (ILF). Scatter in the ILF  can cause selection bias where sources detected at distance   have a greater median luminosity than those detected close by. Median luminosity divides any given population into equal halves. When the functional form of a distribution is unknown, it can be a more robust diagnostic than those that use trial functional forms.

{\bf Method:} In this work, we employ a statistical  test based on median luminosity and use it to test a class of models for GRBs. We assume that the GRB jet has a finite opening angle and that the orientation of the GRB jet is random relative to the observer. We calculate $L_{median}$ as a function of redshift by simulating GRBs  empirically and theoretically, and use the luminosity-versus-redshift {\it Swift} data in order to compare the theoretical results with the observed ones. 
 The method accounts for the fact that there may be GRBs that go undetected  at some redshifts. 
 
 {\bf Results:} We find that $L_{median}$ is extremely insensitive to the on-axis (i.e. maximal)  luminosity of the jet. }


\maketitle


\section{Introduction}

Gamma-ray bursts (GRBs) are among the most energetic objects in the Universe. 
After their discovery (Klebesadel et al. 1973), it was
shown that there are two types of GRBs (Mazets et al. 1981): short and hard, and long and soft. This is
well supported by  both observational data and statistical analysis
(for a survey and the relevant references see, e.g., Meszaros 2006). 

An intrinsic luminosity function (ILF) is among the most sought-after quantities for any class of astrophysical objects. The intrinsic luminosity function  (ILF) of GRBs has been the subject of many papers in the past, but this subject has been plagued by ignorance of what its physical significance would be even if it were known. In particular, as GRBs are known to be beamed along a jet axis,  the intrinsic luminosity function also sheds light on the angular emission profile of the GRBs, assuming the observers are distributed isotropically.

There exists inevitable scatter in the intrinsic luminosities of GRBs, and it is not clear whether or not this scatter is due to relativistic beaming, as offset viewers may see diminished luminosity. Scatter in the intrinsic luminosity can cause selection bias. When the scatter in luminosity is strong, it is important to quantify the effect in order to distinguish it from true dynamical evolution. Therefore, it is important to study luminosity selection effects in GRBs in order to identify the bias. 

Several attempts have been made in the past to find the ILF by fitting the peak flux distribution and the redshift distribution (Tsvetkova et al 2017, Petrosian et al 2015, Pescalli et al 2016, Guetta and Piran 2006). A common result from all these studies is that the ILF that best fits the data is a broken power law with three free parameters (the low and high power-law index and the luminosity break). Guetta et al. 2005  also tried to get information on the angular distribution of the GRBs by assuming that, in the case of a uniform jet, the quantity $L\theta^2$ may be considered constant. This assumption was implied by the observation on the sample considered by Bloom et al 2003.
 
The question, related to finding the ILF, has even greater significance when considering the ILF of `low-luminosity' GRBs (i.e. $L\leq 10^{49}$ erg/s). Are they low luminosity because they are viewed from far away from the direction of motion of  ---and hence the direction of beaming by---   the material with which the observed photons last interacted, or because material beaming mostly in the direction of the  observer is less luminous? 
Guetta and Della Valle 2007  find that there is a difference in the frequency of occurrence of high-luminosity GRBs and  low-luminosity GRBs. This could be interpreted as the existence of two physically distinct classes of GRBs (Cobb et al. 2006, Soderberg et al. 2006b) and whereby low-luminosity
GRBs are (intrinsically) more frequent events than high-luminosity GRBs.
Alternatively, there could simply be a single population whose rate is luminosity dependent.
This single population originates in both the isotropic and subenergetic component, detectable only in nearby GRBs, and the highly collimated component, observable by sampling huge
volumes of space and thus mainly detectable in high-z GRBs.

In this paper, we propose a physical hypothesis for the ILF. We test this physical hypothesis against observed data using statistical tools based on median luminosity where the best-fit model is the one that divides the data most symmetrically. 

The diagnostic we suggest is based on comparison of  the {median} luminosity $L_{median}$ of a set of GRBs that have a known intrinsic luminosity function with the observed luminosity of a wide range of GRBs. The advantage of this method is that it  requires only a small data set.

Though we perform the statistical analysis for a large GRB sample covering both low and high redshifts, the diagnostic proposed in this paper is well suited even for small sets of GRBs, such as  low-redshift GRBs, $z \le 0.15$, of which  only about $10 $ have been discovered, as long as the set is not chosen with luminosity bias.

Several authors have shown that the luminosity function evolves significantly with redshift (Petrosian et al 2015, Pescalli et al 2016, Lloyd-Ronning et al 2019). However, the method proposed in this paper is independent of the redshift distribution. Moreover, we show that it is also independent of the maximum luminosity of GRBs.
The paper is organized as follows:   
In section 2, we discuss our choice of mathematical form of luminosity function, and its physical motivation.   
In section 3, we outline how to calculate the median  bolometric luminosity $L_{median}$ of detected sources as a function of redshift  given some intrinsic luminosity for all of them.  In section 3,  we compare $L_{median}$ calculated from  a set of possibilities, and show that the data are consistent with our physical hypotheses. Finally, in Section 4 we report our conclusion. 

 \section{The physical model and implications for the form of the GRB ILF}
 
 We consider a model of a GRB jet with constant intrinsic
luminosity, beaming angle $\theta_0$, and bulk Lorentz factor $\Gamma$. We are aware that these assumptions are questionable. There have been many papers (Shuang-Xi et al 2017, Ackermann et al 2011, Perna and Loeb 1998, Zou et al 2017, Ghirlanda et al 2013, Tang et al 2015)
dealing with beaming angle and bulk Lorentz factor measurements showing a broad range of values.

In this paper we consider the possible ILF forms described above and test these models against observed {flux} distributions. In order to constrain our model parameters, we use the Fermi GBM and Swift sample of GRBs described in Sect. 2, thereby finding the physical model that explains the data most accurately.
In these works, different values of the jet Lorentz factor and beaming angle are considered in order to fit the observed data. In the present paper, we want to test a model where the jets have the same lorentz factor and opening angle and the difference in the observation is due to the offset of the observers and/or to the absorption material along the line of sight.  We assume that some fraction of GRBs are observed head-on by observers within the opening angle of the jet ($\theta \le \theta_0$), while another set of  offset observers view the jet from outside the  jet opening angle ($\theta \ge \theta_0$). In the second set, the offset observers, we can distinguish  between observers just beyond the edge of the jet ($\theta-\theta_0 \ll \theta_0$) and those far off-axis ($\theta -\theta_0 \gg \theta_0$). Assuming there is no selection bias in favor of larger luminosities, the idea behind the former assumption is that half of all GRBs within a given data set should lie between approximately  the $25^{th}$ and $75^{th}$ percentile of the luminosity distribution at the redshift of the GRB, and only $25\%$ and $10\%$ should lie below the $25^{th}$ and the $10^{th}$ percentile respectively. Therefore, for example, if a sample of four GRBs all lie below the $L_{median}$ predicted by a particular hypothesis, we can rule out the  hypothesis  with a confidence of $15/16$ or  $\sim 94$. If  we can compute the value of $L_{median}$ corresponding to the $25^{th}$ percentile predicted by a given hypothesis, and all four GRBs lie below this value, then  we may rule out the hypothesis to a confidence of $(4^4  -1)/4^4$ or  $\sim 99.6$ \%, and so forth. In general, the probability that in a sample of $N$ GRBs all will lie below the $j^{th}$ percentile is $(j/100)^N(1-j/100)^N$. Therefore, measurement of as few as for example four GRBs at a given redshift would rule out, with reasonable confidence, a median luminosity of more than ten times their observed average luminosity. 

In order to test the model, we first consider a case where the prompt  GRB emission is in a narrow(`pencil') beam, where all the GRBs are identical, and where the distribution of observers is isotropic.
They would nevertheless be distributed in apparent luminosities $L$ because different observers would see them at different viewing angles. If the beam can be considered a pencil beam, then the received luminosity is proportional to $1/(\Gamma(1-\beta \cos \theta))^4$, whereas the solid angle within the cone defined  by the viewing angle is $1-\cos \theta$. However, for viewing angles that are much greater than $1/\Gamma$, $1-\cos \theta \simeq 1-\beta \cos \theta \propto L^{1/4}$. Hence the value of $\alpha$ should be $5/4$.

For GRBs observed slightly beyond the edge of the jet, the source should be considered extended. The effective solid angle from the most visible part of the source is then proportional to the viewing angle from the closest part of the source to the observer's line of sight, and one picks up an extra factor, namely the Doppler factor. The luminosity therefore roughly follows $(1-\cos \theta)^3$, and $\alpha$ takes a value of $4/3$. A description of the model is provided by Banerjee et al 2021. Here, we briefly describe our basic assumptions. We distinguish three possible physical scenarios.

CASE I

Optically thick scenario.  The head-on observers do not see any emission as the material in the jet traps the photons, forcing them to exit the jet region, and therefore the jet is optically thick. Observers outside the jet will see different emission according to their position relative to the jet opening angle. As discussed above for viewing angles that are much greater than $1/\Gamma$, $1-\cos \theta \simeq 1-\beta \cos \theta \propto L^{1/4}$. Hence, the value of $\alpha$ should be $5/4$. For GRBs observed slightly beyond the edge of the jet, the source should be considered extended. The effective solid angle from the most visible part of the source is then proportional to the viewing angle from the closest part of the source to the observer's line of sight, and one picks up an extra factor, namely the Doppler factor. The luminosity therefore roughly follows $(1-\cos \theta)^3$, and $\alpha$ takes a value of $4/3$.

CASE II

Optically thin scenario. For head-on observers, the luminosity function for a set of identical GRBs would be approximately a delta function $\delta(L-L_{max})$ if the beam were optically thin.  This case is excluded  from observations because it predicts more bright bursts than are observed. For observers outside the jet, we get the same as CASE I.
 It is important to notice that this case is excluded because we assume that the prompt emission is produced only within the jet opening angle. 
However, it may be possible that both short and long GRB jets
are structured. In this case, given the jet structure, one can explain the apparent luminosity
function of the GRBs by on- and off-axis observers for the optically thin
scenario. The low luminosity of GRB 170817A can be explained very well in this model as a 
prompt emission production above the jet opening angle. It may be that this
GRB was detected off axis (see Granot et al 2018).
The luminosity function in the case of the structured jet model was derived by Pescalli et al. 2015
(see Eq.15 of Pescalli et al 2015). The luminosity function in this case can be represented as a power law with a slope of $\sim 1.25$ (Pescalli et al 2015). We consider this 
luminosity function in our analysis (see Table 1).

CASE III

Partially optically thick scenario: The head-on observers may see emission from an optically thin emission region, where the photons last interacted with a region that is obscured from the observer by high optical depth (Eichler 2014).
When the shell is optically thick, observers outside the jet see the same LF as the observers outside an optically thin shell because they are seeing photons that are scattered backwards in the frame of the shell. However, observers within the opening angle of the jet cannot see photons backscattered from material moving directly at them. Therefore, there is a `blind spot'. These observers see backscattered photons only from jet material outside the blind spot. The calculation of the size of the blind spot, $\theta_{\rm BS}$, was outlined by  Eichler (2014). This latter authors shows that the minimum angle of the blind spot depends on the optical depth of the shell and on the distribution of circumburst material. 
The blind spot means that there is a minimum offset between the observed material and the line of sight. This is in contrast to observers outside the jet, who can come arbitrarily close to the edge of the jet.  Observers within the jet see a light echo from those parts of the jet that are outside the blind spot. Most of the time-integrated signal comes from an annular just outside the blind spot and  comparable in solid angle to that of the blind spot itself. If the observer is well within the jet, the annulus completely surrounds the blind spot, but if the blind spot touches the edge of the jet then the annulus is only partial and the observer very close to the edge  of the jet sees an annulus that is only about half of the circle. The geometry of this model is discussed in detail by Vyas, Peers, and Eichler (2021). 
For observers just outside the jet, the luminosity function can be represented by a power law as described above. For observers inside the jet, the luminosity function depends on the distribution of the sizes of blind spots. It is reasonable to represent this latter distribution by another power law. The luminosity function depends on the circumburst material and the illumination of the shell from behind.  The total luminosity function is then the sum of the luminosity function for observers outside the jet and the ones inside the jet. Therefore, we always write the ILF as the sum of two terms, $N(L)=N_{\rm in}(L)+N_{\rm out}(L)$. The $N_{\rm out}(L)$ is a power law as discussed above. If the distribution of the blind spot sizes is scale free then it is reasonable to represent $N(L)_{\rm in}$ as a different power law. In this case, the total ILF can be represented by the sum of two broken power laws:
$N(L) \propto (L/L_0)^{-\alpha}+\epsilon N(L)\propto (L/L_0)^{-\beta}$, which can be also approximated as
a broken power law, $N(L) \propto (L/L_0)^{-\alpha}$ for $L<L_0$ and $N(L)\propto (L/L_0)^{-\beta}$ for $L>L_0$.

  \section{Median luminosity for gamma-ray bursts}
 \label{sec1}
In this section we  determine the median luminosity for a given luminosity distribution as a function of the cosmological redshift.
In order to evaluate the median luminosity, we use the relation
\begin{equation}
\int_{L_{min}}^{L_{median}}N(L)dL=\int_{L_{median}}^{L_{max}}N(L)dL
\label{m}
,\end{equation}
such that $L_{median}$ is the luminosity dividing the curve into two equal parts. Here, $N(L,z)$ is the luminosity distribution function of a class of emitters, or in other words is the rate of detected GRBs per unit comoving volume at a redshift $z$ between $L_{min}$ and $L_{max}$.

 We consider two possible forms of the ILF:
The first is the power-law ILF form, $N(L)dL \propto L^{-\alpha}$, which represents the physical case where all the observes are outside the jet $\theta>\theta_0$.
 We then calculate the median luminosity $ L_{median}(z, \alpha)$ as a function of redshift z.
 
 We show that  the data for  nearby, low-luminosity  GRBs and distant, high-luminosity  GRBs are best fitted by a value of $\alpha$ close to $4/3$, and that $L_{median}(z)$  increases with distance by many orders of magnitude within the measured range.  Thus, the low luminosity of nearby GRBs is entirely consistent with the universality of GRB energies, and most of the variation in $E_{iso}$ is attributable to the variation of viewing  angles among the observers, who, according to hypothesis, see an ILF index $\alpha$ ranging from $5/4$ to $4/3$.  We note that this hypothesis (Eichler and Levinson 2004, Eichler and Levinson 2006, Levinson and Eichler 2005) neatly explains both the Amati relation (Amati et al 2002) and the Ghirlanda relation (Ghirlanda et al 2004).  

We consider also a broken power law ILF that represents both the observers inside and outside the jet as explained in the previous section:
\begin{equation}
N(L)\propto (L/L_0)^{-\alpha,-\beta}
\label{broken}
.\end{equation} 
Several authors have checked this ILF form and find the best fit values for the free parameters
by comparing the expected  logN-logS with the observed one (Pescalli et al 2016, Guetta and Piran 2007). Most of the authors agree that the parameters should be in the range of $\alpha\sim 1.3$ (for $L<L_0$) and $\beta\sim 1.8$ (for $L>L_0$)
and a break luminosity $L_0=10^{51}$ erg/sec.
We tested other possible values of the free parameters using the median luminosity method.

The local luminosity function of GRB peak luminosities $L$, defined as the comoving space density of GRBs in
the interval $L$ to $L + dL$, is
\begin{equation}
{\cal N}(L)=c_0 \Big( \frac{L}{L_{max}}\Big)^{-\alpha}
,\end{equation}
where $c_0$ is a normalization constant with a dimension of $1/L,$ meaning that the integral over the luminosity function is equal to unity.

 In our analysis, we consider the evolution of the luminosity function with redshift ${\mathbf{z}}$ including the k-correction effects, as were considered in our earlier paper (Banerjee et al 2021). We consider $N(L,z)=N(L)N(z)$ (as in (Lloyd-Ronning et al 2002, Yonetoku et al 2004)), with $N(z)$ defined as (Lloyd-Ronning et al 2002, Petrosian et al 2015)
\begin{equation}\label{Rcj}
N(z)=\rho_o\rho(z),
\end{equation}
where the star formation rate is given by (Madau 2014)
\begin{equation}
\rho(z)=\frac{(1+z)^{2.7}}{1+[(1+z)/2.9]^{5.6}}.
\end{equation}
 Here,  $\rho_0$ is the GRB rate at $z=0$ expressed in units of $Gpc^{-3}year^{-1}$.
 
Therefore, the evolution of the luminosity function with redshift is given by 
\begin{equation}
N(L,z)=\rho_0c_0\Big( \frac{L}{L_{max}}\Big)^{-\alpha}\frac{(1+z)^{2.7}}{1+[(1+z)/2.9]^{5.6}}
\label{1}
\end{equation}
out to some maximum redshift $z_{max}$, beyond which $N$ is assumed to vanish.

However, for the main focus of this paper, namely the dependence
of mean luminosity on redshift,
these purely redshift dependence aspects are irrelevant and can be omitted. This can be seen if we substitute the form of $N(L,z)$ from Eq. \ref{1} into Eq. \ref{m}. The purely z-dependent terms get cancelled out from right hand side(r.h.s) and left hand side(l.h.s), leaving behind the $z$ dependence coming from $L_{min}$ as shown below.

 Now using Eq. \ref{1} in Eq. \ref{m} and solving for $L_{median}$ we get,
\begin{equation}
L_{median}(z)=\left( \frac{L_{max}^{-\alpha+1}+(L_{min})^{-\alpha+1}}{2}\right)^{\frac{1}{-\alpha+1}}
\label{mm}
.\end{equation}
Similarly,
\begin{equation}
L_{25}^{-\alpha+1}- L_{min}^{-\alpha+1} = \frac{(L_{max}^{-\alpha+1}-L_{25}^{-\alpha+1})}{3} ,\end{equation}  
and so 
\begin{equation}
L_{25} =  \Big[\Big(\frac{3}{4}\Big)(L_{max}^{-\alpha+1} + L_{min}^{-\alpha+1})\Big]^{1/(1-\alpha)} ,\end{equation}
and the luminosity that established the $j^{th}$ percentile is given by 
\begin{equation}
L_j=  \frac{L_{max}^{-\alpha+1}+
 L_{min}^{-\alpha+1}}{(1-j)(1-\alpha)}
 .\end{equation}
In order to generalize the above expression for any given cosmological model, let us now consider that $L_{min}$ is established by the detector flux, whereby
\begin{equation}
L_{min}=4\pi F_{th}D_L^2
\label{3}
.\end{equation}
Here, $D_L$ is the luminosity distance, which is a function of the redshift $z,$ and $F_{th}$ is the threshold flux. Therefore, substituting $L_{min}$ into Eq. \ref{mm}, we get
\begin{equation}
L_{median}(z)=\left( \frac{L_{max}^{-\alpha+1}+(4 \pi D_{L}^2(z)F_{th})^{-\alpha+1}}{2}\right)^{\frac{1}{-\alpha+1}}
\label{mmm}
.\end{equation}
In order to obtain the nature of the variation of $L_{median}$, let us consider the $\Lambda$cold dark matter (CDM) model.
The expression for the luminosity distance in terms of the redshift $z$ for the $\Lambda$CDM model is given by (assuming the value of the scale factor at the present time is normalized to $1$)
\begin{equation}
D_L=\frac{c}{H_0}(1+z)\int_0^{z'}\frac{1}{((1+z')^3\Omega_{m_0}+\Omega_{\Lambda_0})^{1/2}}dz'
\label{dl}
,\end{equation}
where $\Omega_{m_0},\ \Omega_{\Lambda_0}$, and $H_0$ are the present values of the matter density, the cosmological constant, and the Hubble parameter, respectively.
\subsection{Comparing with observed data}
We now use  Eq. \ref{dl} in Eq. \ref{mmm} and get the variation of $L_{median}$ with $z$. However, since  the integral can be solved only numerically for $\Lambda$CDM, we can  only  get a numerical solution for the median luminosity.

\begin{figure}[htbp!]
\centering
\subfigure[]{
    \includegraphics[width=\columnwidth,height=9cm]{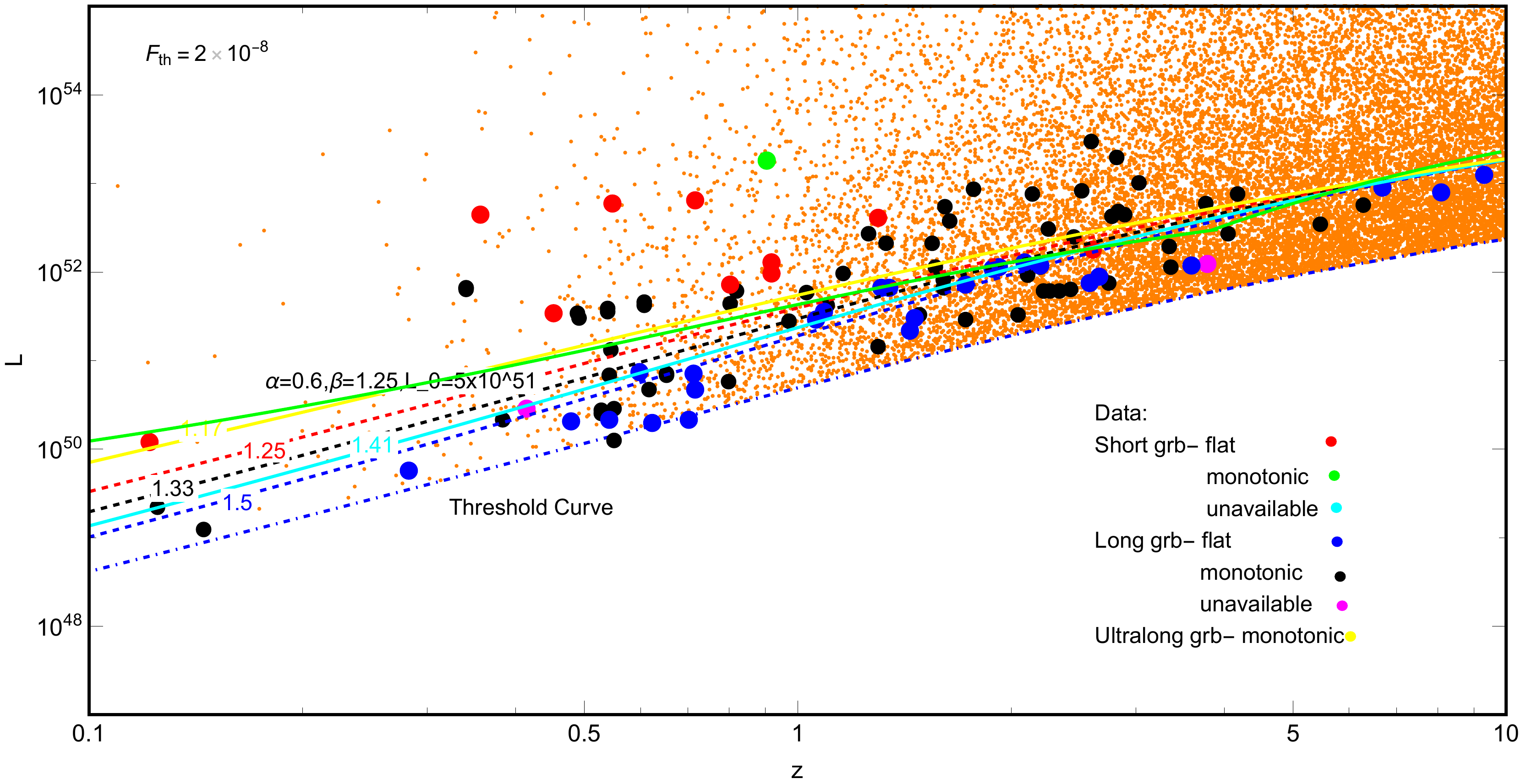}
}
\subfigure[]{
    \includegraphics[width=\columnwidth,height=9cm] {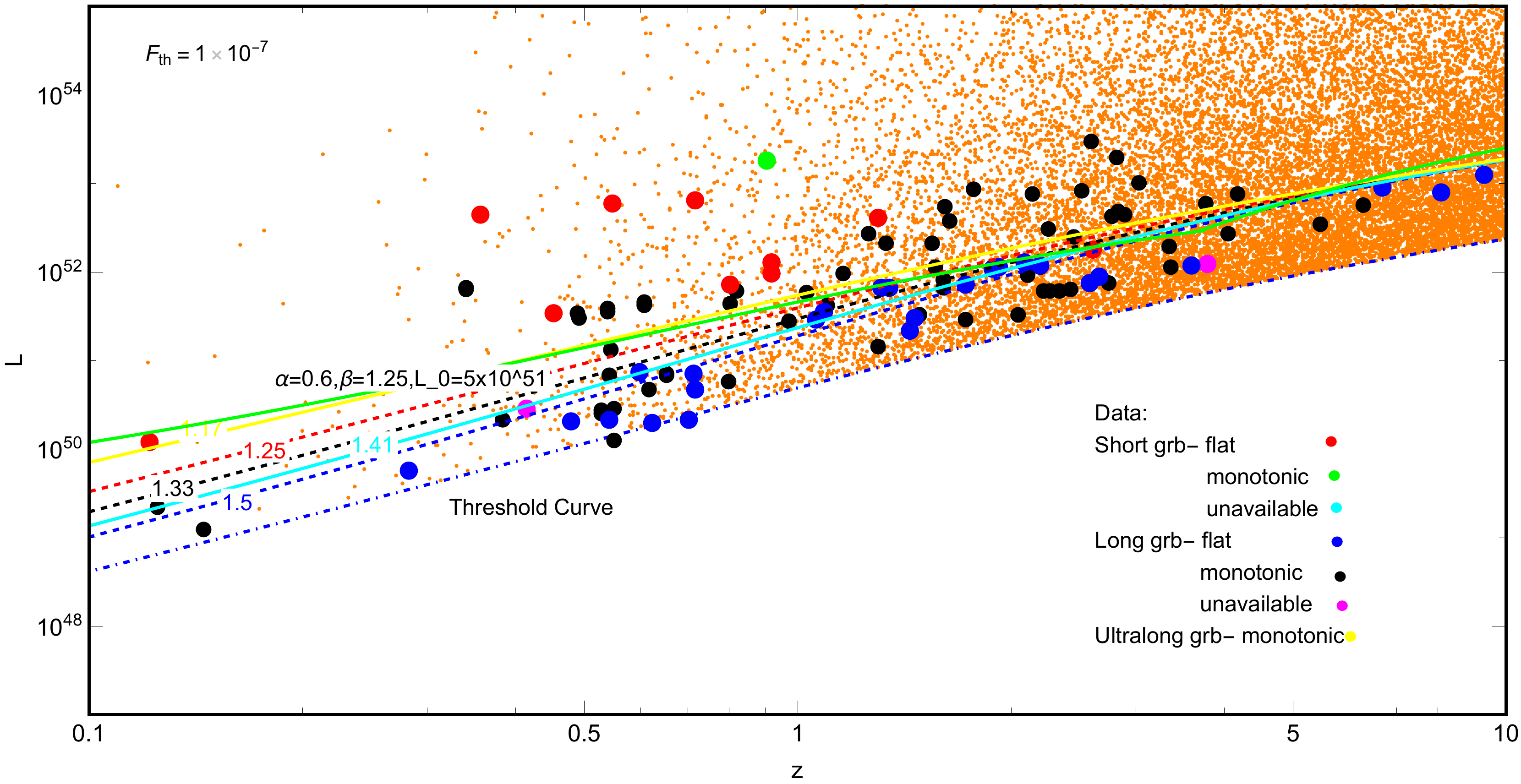}
}
\caption{Plot showing the variation of $L_{median}$ (in erg/s) with $z$ along with observed {\it Swift} data and simulated data. The plots in panel (a) correspond to $F_{th}=2\times 10^{-8}$ erg/cm$^2$/s. The plots in panel (b) correspond to $F_{th}=1\times 10^{-7}$ erg/cm$^2$/s. The values of the parameters are marked on the plots. The orange dots represent the simulated data. The other, larger dots correspond to the observed data for long, short, and ultralong GRBs (Goldstein et al 2017, Cano et al 2017, MichałowskI et al 2018, Gehrels et al 2006, D’Avanzo et al 2014, Demianski et al 2017). The terms monotonic, flat, and unavailable correspond to the nature of afterglow data for the corresponding GRB. The curve marked threshold represents the threshold value (for the rest of the plots in this paper) of $L$ at a given $z$. For comparison, we also plot the curve (green and brown) corresponding to the ILF given by a broken power law.}
\label{sim1}
\end{figure}
\begin{figure}[htbp!]
\centering
\subfigure[]{
    \includegraphics[width=\columnwidth,height=9cm]{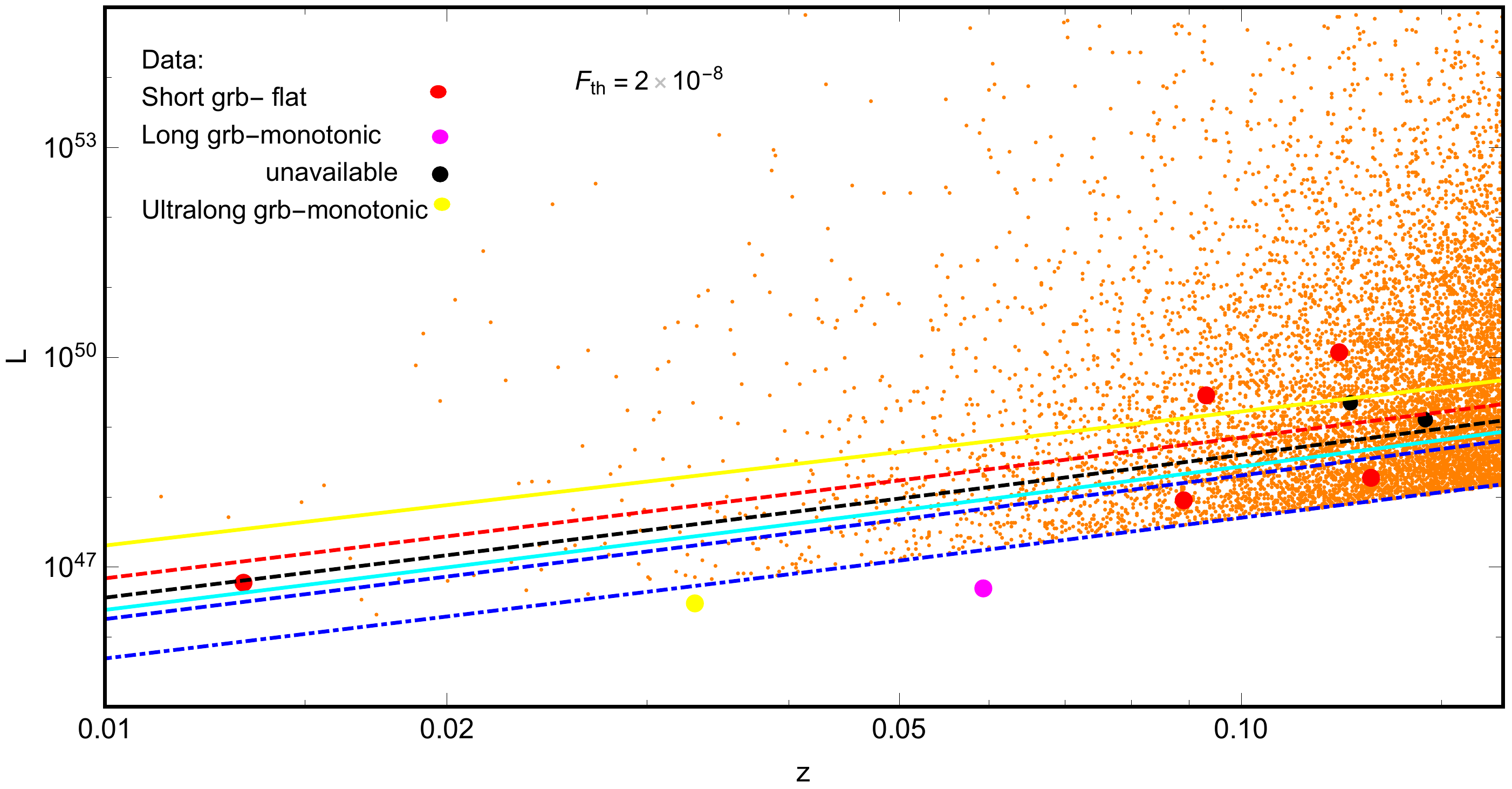}
}
\subfigure[]{
    \includegraphics[width=\columnwidth,height=9cm] {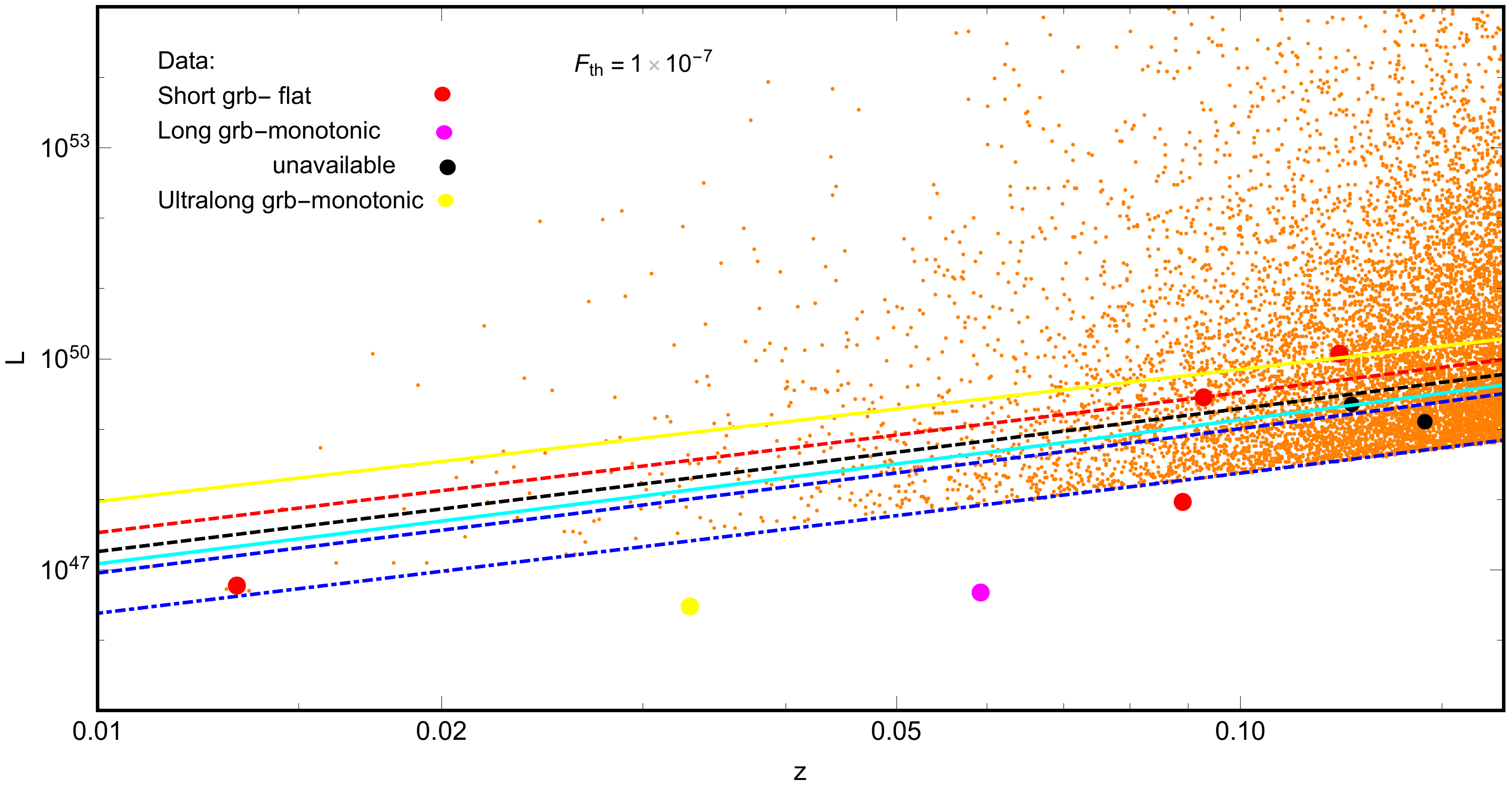}
}
\caption{ Variation of $L_{median}$ (in erg/s) with $z$  for small redshifts and comparison with simulated and observed data. $F_{th}$ is in units of erg/cm$^2$/s. The plots in the panel (a) correspond to $F_{th}=2\times 10^{-8}$. The plots in panel (b) correspond to $F_{th}=1\times 10^{-7}$. The values of the spectral index $\alpha$  are marked on each curve. The orange dots are for simulated data. The larger dots correspond to the observed data for long, short, and ultralong GRBs (Goldstein et al 2017, Cano et al 2017, MichałowskI et al 2018, Gehrels et al 2006, D’Avanzo et al 2014, Demianski et al 2017). The green dots are for short GRBs with monotonic afterglow, magenta dots for long GRBs with unavailable afterglow data, and the yellow dots correspond to ultralong GRBs with monotonic afterglow data. }
\label{sim2}
\end{figure}

{\it Observed Data Sample}: Most GRBs do not have known redshifts, and  knowledge of the redshift, which generally requires identification of a host galaxy, is probably biased towards the high-luminosity GRBs, which receive more attention, and for which it is easier to find a host galaxy (the automatic slewing capabilities of Swift lessen this bias, but do not eliminate it).
However,  for very nearby GRBs ($z \lesssim 0.15)$, a supernova is generally observed, and so no matter how dim the GRB is, its coincidence in time and space with the supernova ensures the determination of its redshift. If there is no associated supernova, coincidence with a nearby Galaxy (which is more statistically significant than coincidence with a distant Galaxy), can reveal the redshift. The discovery of a kilonova or a gravitational wave signal can, as in the case of GRB 170817 (LIGO 2017, Goldstein et al 2017, Savchenko et al 2017) help in confirming the host galaxy. LIGO and VIRGO observations can shed light on short GRBs, but not the mostly long GRBs that we have considered for our analysis. 
In our present work, the observed data are taken from Goldstein et al. 2017, Cano et al. 2017, MichałowskI et al. 2018, Gehrels et al. 2006, D’Avanzo et al. 2014, and Demianski et al. 2017). Our sample consists of data collected from different sources in such a way that it spans a wide range of redshift starting from the nearest GRB ($z=0.00867$) up to redshifts as high as $9.$ 
The low-redhsift GRBs are mostly associated with supernovae. The sample has also been chosen such that we have adequate data  (including the values of spectral peak, redshift, and flux) for each GRB so that they can be included in all our plots. The Swift data comprise 17 short GRBs, 115 long GRBs, and 3 ultra-long GRBs.

We further divided the data sets based on their afterglow nature: flat or monotonic. 

 In Banerjee et al. (2021), we obtained the best-fit parameter values for single and broken power-law luminosity functions using peak flux distribution data  for Swift and Fermi GRBs. The Fermi GRB sample consisted of 918 long GRBs obtained from Gruber et al. (2014), Kienlin et al. (2014), Bhat et al. (2016), and Meszaros and Meszaros (1996). The Swift sample on the other hand consisted of $365$ long GRBs obtained from the Swift catalogue, Gruber et al. (2011), and Gruber (2012).
For both cases, we obtained the histogram for the differential source counts using their peak flux distributions and compared it to the peak flux of the whole sample with known and unknown redshifts. 
Using the method of $\chi^2$ minimization and p value, we obtained best-fit parameter values of (i) $\alpha=1.17$    for single power law and (ii)  $\alpha=0.6_{-1}^{+1},\ \beta=1.25_{-0.05}^{+0.05}$, and $L_0=5_{-2}^{+2}\times 10^{51}$ for
a broken power law.

In Fig. \ref{sim1}, we plotted the variation of $L_{median}$ with redshift for $\alpha=1.5,\ 1.33,\ 1.25$,\ $1.41$, and $1.17$, and $L_{max}=3\times 10^{53}$ for two different values of threshold flux $F_{th}=2\times 10^{-8}$  and $F_{th}=1\times 10^{-7}$. We also show the results for $L_{median}$ corresponding to the ILF given by a broken power law defined in Eq. \ref{broken} using the best-fit parameter values obtained from our analysis of the $LogN-LogF$ plot in Banerjee et al. (2021). Here, we show that the median luminosity can also be used to check whether or not these models are accurate. Henceforth, $L$ and $F$ correspond to the peak bolometric luminosity and flux, respectively, with $L$ in units of erg/s and $F_{th}$ in units of erg/cm$^2$/s.

In Fig. \ref{sim2}, we highlight the low-$z$ region.
We also show the position of GRB170817 (Goldstein et al 2017) with $L_{peak}=1.3\times 10^{47}$ ergs/s and $z=0.009$ in our plot.
In both figures, the values of $\alpha$ are mentioned in the plot as well.

 As we are plotting $L_{median}$, the curve that best fits the data should have an equal number of data points below and above the corresponding theoretical curve. Counting the number of data points above and below each curve for a single power law in Fig. \ref{sim1} we find that for $F_{th}=2\times 10^{-8}$, the curve corresponding to $\alpha=1.17$ is the most symmetrical one, with $\alpha=1.25$ lying close by. The curves corresponding to other values of $\alpha$ are much less symmetrical and can therefore be rejected. We find that if we increase the value of $F_{th}$ by a factor of five as shown in Fig. \ref{sim1}(b), the most symmetrically curve for a single power law is now given by a higher value of $\alpha=1.41$ with $\alpha=1.33$ lying close by. 
 
 For the broken power law, we plot the best-fit values obtained in  Banerjee et al. (2021): $\alpha=0.6,\ \beta=1.25,$ and $L_0=5\times 10^{51}$. We find that a broken power law with these parameter values fits the data most closely for $F_{th}=2\times 10^{-8}$. For $F_{th}=1 \times 10^{-7}$, we obtain the same results.
The exact numbers of the counts are shown in Tables \ref{pvalue1} and \ref{pvalue2}.
 
 For very low redshifts, as in Fig. \ref{sim2}, the broken power law reduces to a single power law case, depending only on the spectral index $\alpha$. Hence, we do not show it in Fig. \ref{sim2} as it is not relevant.

For the present case, in order to quantify the goodness of the fit using the present sample, we compare the theoretical and observed results to obtain the p values for single and broken power laws using the best-fit parameter values obtained in Banerjee et al. (2021). For this, we divided the analysis into ten redshift bins, and compared the median luminosity from the observed data and the theoretical median luminosity for each redshift bin. In Table \ref{table}, we show the p values obtained using the best-fit parameter values for both single and broken power law cases. 

We note that we also plot the threshold curve in all the figures for reference. Some of the data points lie either on or slightly below it. However, these GRBs are either very long, have uncertainties in their redshifts, or are affected by X-ray flash. Hence, all of them have properties different from other GRBs.
From Fig. \ref{sim1}(b), which contains the Swift data, we see that, among the long GRBs, the ones with flat afterglows are much less scattered compared to those with monotonic afterglows. The probability of this being by chance is $0.03$ or $3\%$ at $95\%$ confidence. Thus, this difference is statistically significant.

\begin{table*}[t]
\begin{tabular}{ |c|c|c|c|c| } 
\hline
$F_{th}$ & $\alpha$ & $L_{max}$ 
& Observed points(111) & Simulated points(20000) \\
\hline
$1 \times 10^{-7}$ & $1.33$ & $3\times 10^{53}$ 
& $49$ & $10942$ \\
\hline
$1\times 10^{-7}$ & $1.25$ & $3\times 10^{53}$ 
& $44$ & $9132$ \\ 
\hline
$1\times 10^{-7}$ & $1.17$ & $3\times 10^{53}$  
& $51$ & $10511$ \\ 
\hline
$1 \times 10^{-7}$ & $1.41$ & $3\times 10^{53}$ 
 & $57$ & $11470$\\ 
\hline
$1\times 10^{-7}$ & $1.5$ & $3\times 10^{53}$ 
 & $64$ & $12777$ \\ 
\hline
$1 \times 10^{-7}$ & $\alpha=0.6,\ \beta= 1.25,\ L_0=5 \times 10^{51}$ & $3\times 10^{53}$ 
 & $56$ & $10810$ \\ 
 \hline
\end{tabular}
\caption{Table containing the number of observed (column 4) and simulated (column 5) data points corresponding to $F_{th}=1\times 10^{-7}$. In columns $4$ and $5$, the number in brackets corresponds to the total number of points.}
\label{pvalue1}\end{table*}

\begin{table*}[t]
\begin{tabular}{ |c|c|c|c|c| } 
\hline
$F_{th}$ & $\alpha$ & $L_{max}$ 
& Observed points(135) & Simulated points(20000) \\
\hline
$2 \times 10^{-8}$ & $1.33$ & $3\times 10^{53}$ 
& $102$ & $12990$ \\
\hline
$2 \times 10^{-8}$ & $1.25$ & $3\times 10^{53}$
& $89$ & $11832$ \\ 
\hline
$2 \times 10^{-8}$ & $1.17$ & $3\times 10^{53}$ 
& $74$ & $10608$ \\ 
\hline
$2 \times 10^{-8}$ & $1.41$ & $3\times 10^{53}$ 
 & $106$ & $13926$\\ 
\hline
$2 \times 10^{-8}$ & $1.5$ & $3\times 10^{53}$ 
 & $109$ & $14777$ \\ 
\hline
$2 \times 10^{-8}$ & $\alpha=0.6,\ \beta= 1.25,\ L_0=5 \times 10^{51}$ & $3\times 10^{53}$ 
 & $66$ & $10704$ \\ 
 \hline
\end{tabular}
\caption{Table containing the number of observed (column 4) and simulated (column 5) data points for $F_{th}=2\times 10^{-8}$ corresponding to Swift. In columns $4$ and $5$, the number in brackets corresponds to the total number of points.}
\label{pvalue2}\end{table*}
\begin{table*}[t]
\centering
\begin{tabular}{ |c|c|c| } 
\hline
\quad \quad \quad \quad Parameter Values  \quad \quad \quad \quad  &  \quad \quad \quad \quad $F_{th}$  \quad \quad \quad \quad & \quad \quad \quad \quad p value \quad \quad \quad \quad\\ 
\hline
$1.17$ & $2\times 10^{-8}$ 
 & $0.55$ \\ 
\hline
$1.17$ & $1\times 10^{-7}$ 
 & $0.49$ \\ 
 \hline
$\alpha=0.6_{-1}^{+1}$,\ $\beta=1.25_{-0.05}^{+0.05}$,\ $L_0=5_{-2}^{+2}\times 10^{51}$ & $2\times 10^{-8}$  & 
  $0.70$ \\ 
\hline
$\alpha=0.6_{-1}^{+1}$,\ $\beta=1.25_{-0.05}^{+0.05}$,\ $L_0=5_{-2}^{+2}\times 10^{51}$ & $1\times 10^{-7}$  & 
  $0.66$ \\ 
\hline
\end{tabular}
\caption{Table containing the p values for the single and broken power laws considered in the paper obtained comparing the predictions with the data using the best-fit parameter values obtained by Banerjee et al. (2021).}
\label{table}
\end{table*}
\begin{figure}[htbp!]
\centering
\subfigure[]{
    \includegraphics[width=\columnwidth,height=9cm]{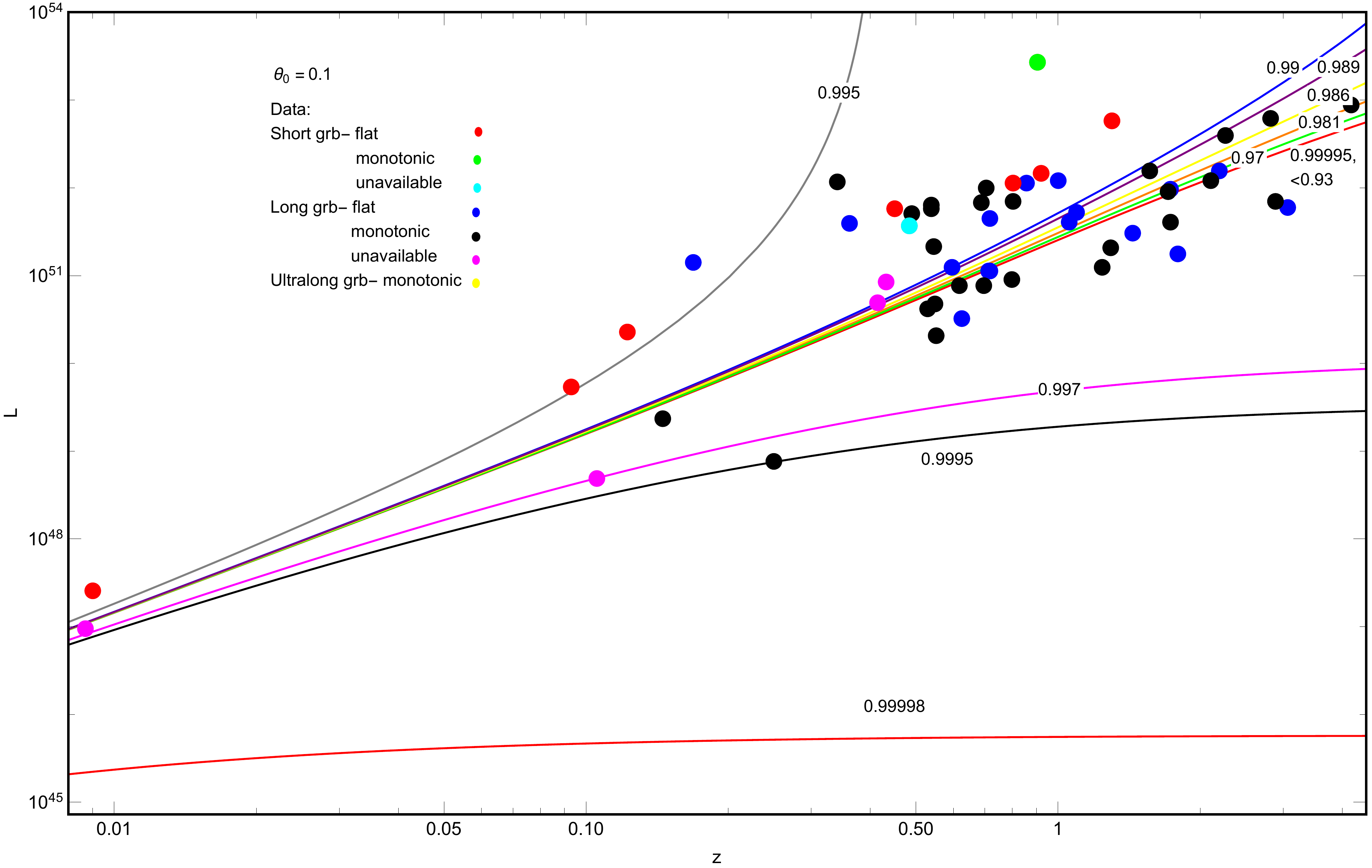}
}
\subfigure[]{
    \includegraphics[width=\columnwidth,height=9cm]{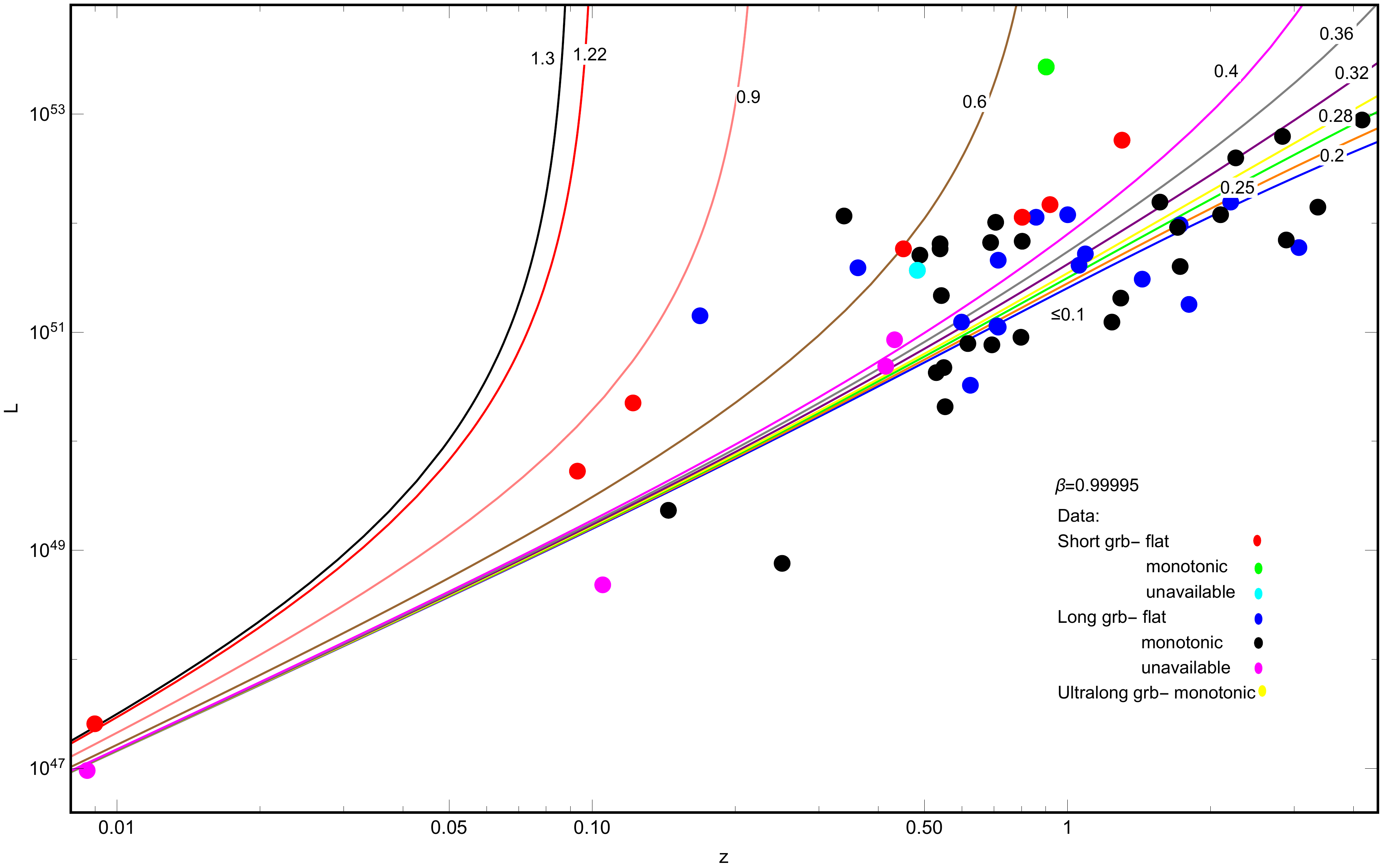}
}
\caption{Variation of $L_{median}$ with $z$ for an extended jet along with Fermi GBM data. The plots in panelg (a) are for $\theta=0.1$. Different curves correspond to different values of $\beta$. Curves in panel (b) are for $\beta=0.99995$.}
\label{theta}
\end{figure}
\begin{figure}[htbp!]
\centering
    \includegraphics[width=\columnwidth,height=9cm]{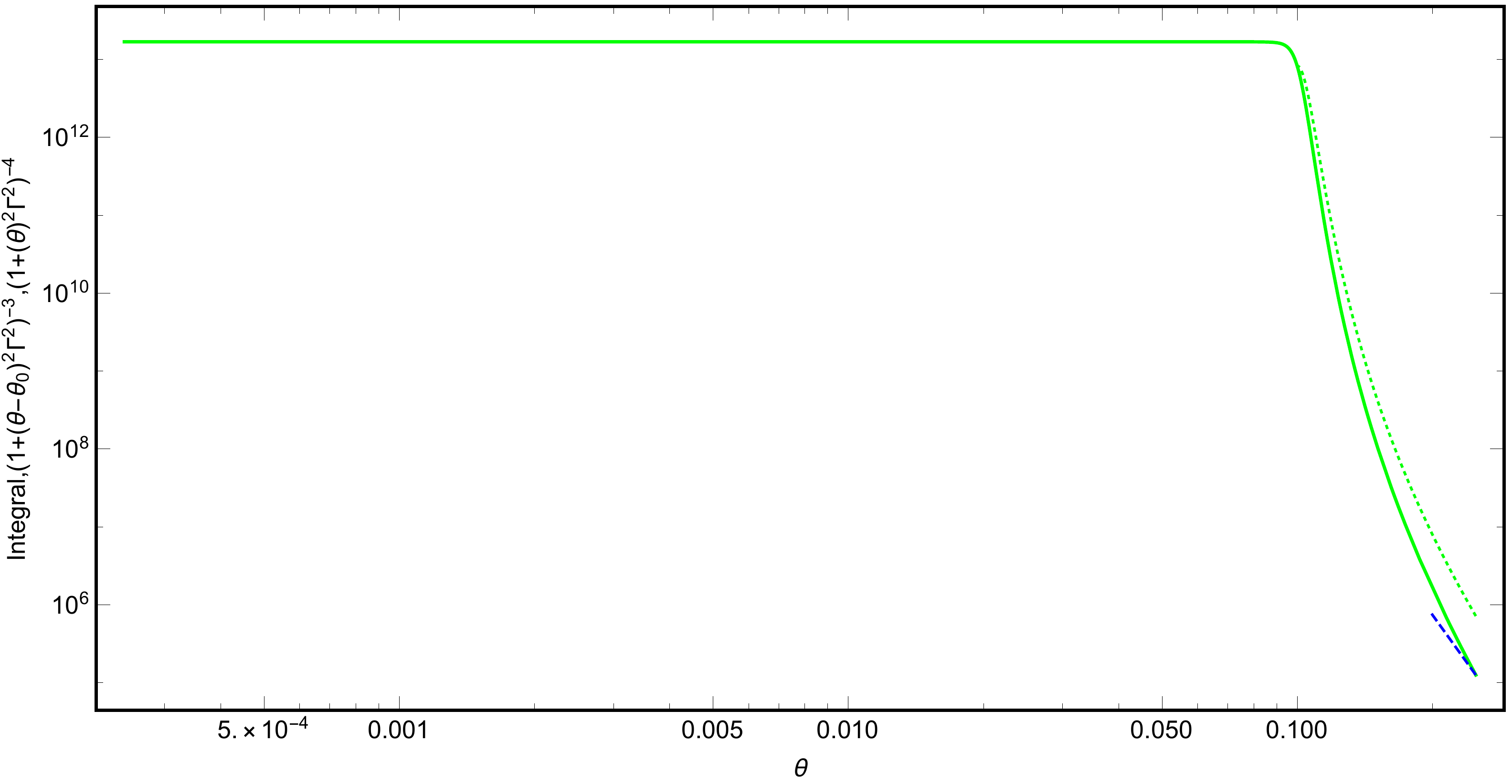}
\caption{Numerical solution of Eq. \ref{extjet} compared to analytical approximate expressions: $[1+(\theta-\theta_o)^2(\Gamma)]^{-3}$ (in the region just beyond $\theta_0$) represented by the green dotted line and $[1+(\theta)^2(\Gamma)^2]^{-4}$ (the region $0.2<\theta<0.3$) represented by blue dashed line. }
\label{comparison}
\end{figure}
\begin{figure}[htbp!]
\centering
    \includegraphics[width=\columnwidth,height=9cm]{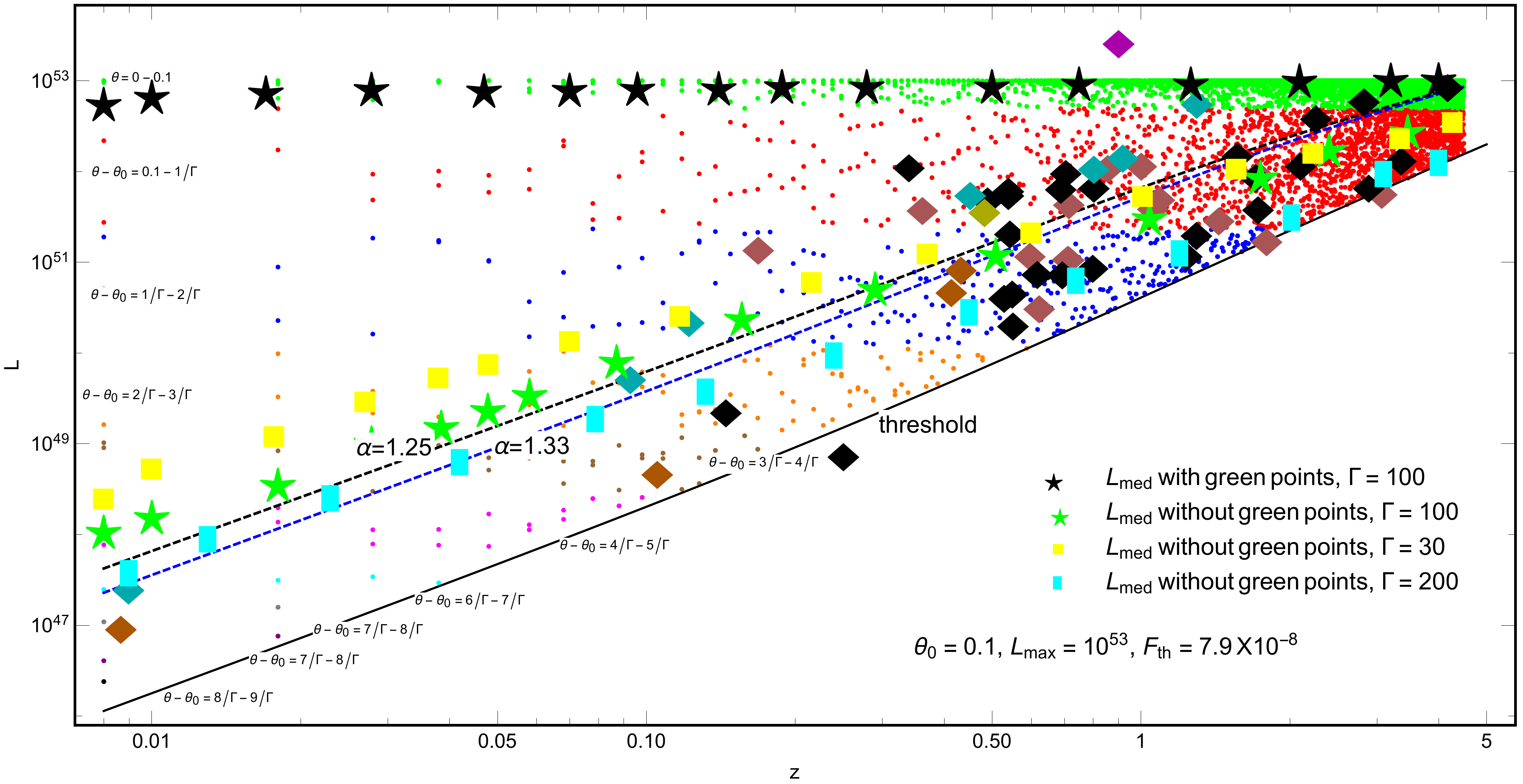}
\caption{Simulated data showing variation of $L$ with $z$ for different ranges of $\theta$. The black stars correspond to the value of $L(z)$ that divides the points equally (above and below). The green stars reflect the same $L(z)$ but without the green points, i.e. for points outside the jet. The yellow squares and cyan rectangles represent similar points ($L(z)$) for $\Gamma=30$  and $ \Gamma=200,$ respectively. The other, larger diamonds correspond to the observed data (Goldstein et al 2017, Cano et al 2017, MichałowskI et al 2018, Gehrels et al 2006, D’Avanzo et al 2014) along with the nature of their afterglows (flat, monotonic, or unavailable): dark cyan = short, flat; dark magenta = flat, monotonic; dark yellow = flat, unavailable; dark pink = long, flat; black = long, monotonic; and dark orange = long, unavailable. The analytical curves for $\alpha=1.25$ (black dot-dashed) and $\alpha=1.33$ (black dashed) are also shown.}
\label{simlowz}
\end{figure}
\begin{figure}[htbp!]
\centering
    \includegraphics[width=\columnwidth,height=9cm]{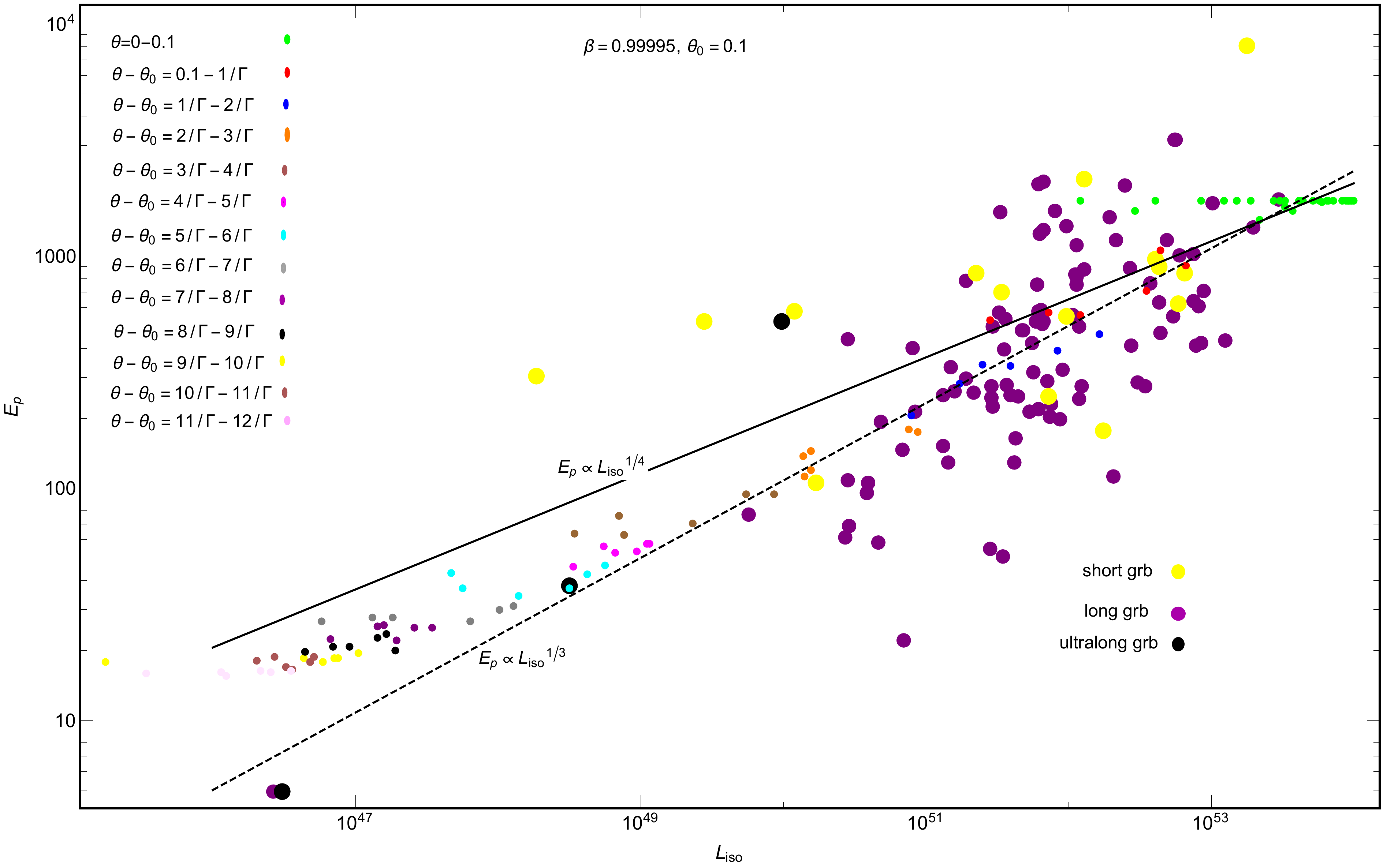}
\caption{$E_p$ vs. $L_{iso}$  as a function of $\theta$ along with observed Fermi GBM data. The ranges of $\theta$ are the same as in Fig. \ref{simlowz}. The lines corresponding to $E_p\propto L_{iso}^{1/4}$ and $E_p\propto L_{iso}^{1/3}$ are also shown.}
\label{yonetoku}
\end{figure}

 We further checked whether or not lowering $L_{max}$ causes any improvement. We find that the change in $L_{median}$ due to a change in $L_{max}$ by a factor of ten is almost negligible. Hence, introducing a spread in $L_{max}$ given by a sum of two luminosity functions will not cause any significant change in $L_{median}$.
 
 As we know, for head-on observers, the ILF for a set of identical GRBs is a delta function $\delta(L-L_{max})$ if the beam is optically thin. Following our earlier analysis, one can easily see that the $L_{median}$ for this case is a constant around $L_{max}$ which clearly does not match the observational data, and is therefore ruled out.

\subsection{Comparing with simulated data}
Observed data may be biased, especially at high redshifts. Hence, in order to improve our statistics, we further compared our theoretical results with simulated data for $L_{median}$ versus $z$.
In order to generate the simulated data, we considered a pencil beam with random orientations of GRBs in the sky. In other words, we assumed that GRBs can occur anywhere within a certain volume at equal cosmic times. We further divided intervals of redshift into equal volumes. As discussed above, the dependence of luminosity on the jet opening angle for a pencil beam is proportional to $1/(\Gamma(1-\beta\cos{\theta})^4$, $\theta$ being the jet opening angle. 

Using this information, we generated the simulated data for $L_{median}$ in terms of redshift with an average of 17 GRBs in each redshift bin($dz=0.01$). The results are shown in Figs. \ref{sim1} and \ref{sim2} for both large and small redshifts for the two threshold values we considered in the earlier section.

In order to compare the theoretical results with simulated data, we once again counted the number of points above and below each theoretical curve and identified the curve that divides the data set most symmetrically. We repeat this for both Figs. \ref{sim1}(a) and (b). 
As above, we find that for $F_{th}=2\times 10^{-8}$, among the chosen values of $\alpha=1.17,\ 1.25,\ 1.33,\ 1.41,\ $ and $1.5$, the $L_{median}$ curves for $\alpha=1.17$ divide the data points most symmetrically with $\alpha=1.25$ lying close by. Higher values of $\alpha$ worsens the fit. 
As we increase the value of $F_{th}$ to $1\times 10^{-7}$, the best-fit curve for a single power law is now given by $\alpha=1.33$. The exact numbers are shown in Tables \ref{pvalue1} and \ref{pvalue2}.

For completeness, using the best-fit values of the free parameters obtained above, we see that they once again divide the simulated points most symmetrically. The exact numbers, for both the thresholds, are shown in Tables \ref{pvalue1} and \ref{pvalue2}.

Thus, comparing our results with both simulated and observed data, we can safely conclude that for the case where the luminosity of the GRBs is given by a power law only, their luminosity distribution with respect to redshift is best given by a pencil beam, that is, $\alpha=1.17$ and $\alpha=1.25$. Higher values of $\alpha$ fit the data very poorly, and are therefore rejected.
This result is strong evidence that the jet may be structured as shown in Pescalli et al. (2015).
The best-fit values of the broken power-law parameters given by Banerjee et al. (2021) give a very good fit to the data for the present case also. 
\subsection{Analytical solution for an extended jet }
Let us now consider an extended jet with an opening angle $\theta_0$ such that $dL_0/d\Omega'=\delta(\theta_0-\theta)$. Here, $dL_0$ is the contribution from the head-on element $d\Omega'$ of the jet, and $\theta$ is the angle between the jet axis (situated at zero) and the observer. Considering the Doppler effect, 
\begin{equation}
\frac{d L}{d\Omega'}=\frac{d L_0}{d\Omega'}\frac{(1-\beta)^4}{(1-\beta\cos{\chi})^4}
,\end{equation}
where $\chi$ is the angle between the observer and the element $d\Omega'$, the total luminosity is obtained by integrating the above expression over $d\Omega'$:
\begin{equation}
    L=\int \frac{d L}{d\Omega'}d\Omega'=\int \frac{d L_0}{d\Omega'}\frac{(1-\beta)^4}{(1-\beta\cos{\chi})^4}d\Omega'
.\end{equation}
Writing $\cos{\chi}=\sin\theta\sin\theta'\cos\phi'+\cos\theta\cos\theta'$, we get
\begin{equation}
    L=\int\delta(\theta_0-\theta)\frac{(1-\beta)^4\sin\theta'd\theta'd\phi'}{(1-\beta(\sin\theta\sin\theta'\cos\phi'+\cos\theta\cos\theta'))^4}
    \label{extjet}
.\end{equation}
The luminosity distribution function is now given by
\begin{equation}
    dN=\frac{\frac{dN}{d\cos\theta}}{\frac{dL}{d\cos\theta}}dL
    \label{extjet1}
.\end{equation}
As we know, $dN/d\cos\theta=constant=K$ where $K$ is the rate of GRBs, which is two per day. 

Using the above ILF, one can now compute the median luminosity, and therefore find the best-fit values of $\Gamma$ and $\theta_0$. For this, we use the Fermi GBM data as before. In Fig. \ref{theta}(a), we plotted the variation of $L_{median}$ with $z$, setting $\theta_0=0.1$. Different curves correspond to different values of $\Gamma$. Plotting the curves along with GBM data, we find from Fig. \ref{theta}(a) that for $\theta_0=0.1$, the best-fit value of $\Gamma=5.15$. $\Gamma=100$ also gives a good fit. However, values of $\Gamma$ such as $10.01,\ 31.6$, and so on give a very poor fit, and $\Gamma$  values of $<2.7$ fit the data quite well. In Fig. \ref{theta}(b), we plot the curves for different values of $\theta_0$ for $\Gamma=100$. Comparing with data, we find that the best-fit value of $\theta_0$ is $0.1$, which is already clear from Fig. \ref{theta}(a). We also see that values of $\theta_0$ of lower than $0.1$ still give a good fit, but the fit becomes poorer above $\theta_0=0.1$.

 In Fig. \ref{comparison}, we compare the numerical solution of Eq. \ref{extjet} to analytical approximate expressions-$[1+(\theta-\theta_o)^2(\Gamma)]^{-3}$  and $[1+(\theta)^2(\Gamma)^2]^{-4}$ .
In Fig. \ref{simlowz}, we show a simulated plot for $L$ weighted by a star formation $z$ dependence $(1+z)^k$ for $k=1.4$ (Lloyd-Ronning et al 2002, Yonetoku et al 2004). This plot was obtained for $\theta_0=0.1,\ \Gamma=100, $ and$ \ L_{max}=10^{53}$. Points are above the Fermi GBM threshold are shown. We also show the threshold curve and analytical curves for $\alpha=1.33 $ and $ \ 1.25$. As we can see, the contribution of jet elements outside the opening angle ---that is, larger values of $\theta-\theta_0$--- becomes important and observable only near low $z$. For higher redshifts, the main contribution comes from elements near to the jet axis, that is, in and around the opening angle.
We also show the real $L_{median}$ values that divide the number of GRBs equally at a given $z$. In order to clearly see the effect of including the GRBs within the jet, we show the real $L_{median}$ for both with and without the GRBs inside the jet for $\Gamma=100$. As we can see, when we include the GRBs within the jet, the $L_{median}$ curve no longer passes through the data point, and therefore gives a poor fit. However, for highly opaque medium, that is, where we can see only those parts of the jet where the velocity vector is sufficiently misaligned with the line of sight, the $L_{median}$ curve gives a much better fit. For reference, we also show $L_{median}$ for $\Gamma=30,\ \Gamma=200$.

Finally, in Fig. \ref{yonetoku}, we plot the $E_p$ versus $L_{iso}$, varying both with respect to $\theta$ with a spread in $L_{max}\sim \mathcal{O}(10)$. From Eichler and Levinson (2006) and Ioka and Nakamura (2001), we know that the maximum frequency goes as $\nu_{max}\propto 1/(1-\beta\cos\chi)$. Therefore, expressing $\chi$ in terms of $\theta$ and integrating over the jet element, we get $E_p$ as a function of $\theta$. Figure \ref{yonetoku} was obtained for $\theta_0=0.1$ and $\Gamma=100$. We plot $E_p\propto L_{iso}^{1/4}$ and $E_p\propto L_{iso}^{1/3}$ curves for reference. As we can see, for larger angles (away from the edge), the simulated plot follows the curve $E_p\propto L_{iso}^{1/4}$  and then near the edge it goes more like $E_p\propto L_{iso}^{1/3}$.

\section{Conclusions}
\label{sec5}

In this paper, we study the behaviour of the quantity $L_{median}$ for a given luminosity function. We studied the behaviour for the luminosity functions given by a power law and we find that, under very general assumptions, $L_{median}$  always increases with redshift and therefore distance, as expected. We also find it to be very insensitive to $L_{max}$, the intrinsic luminosity seen by an on-axis observer.
 In our present work, we have chosen a class of physical models described by Eichler (2014; and references therein). This model is essentially a shell with baryonic material illuminated from behind that is possibly optically thick. For simplicity, we consider a constant bulk Lorentz factor and an opening angle $\theta_0$.
The GRB ILF consists of a contribution of $N_{\rm out}(L)$ observers outside the jet and $N_{\rm in}(L)$ inside $\theta_0$.
 We calculated $N_{\rm out}(L)$ and show that it can be approximated as $N(L)dL \propto L^{-\alpha}dL$ where $\alpha=4/3$ for observers just outside the jet and $\alpha=5/4$ for observers far from the jet. This power law extends to a maximum value that corresponds to the observers within $1/\Gamma$ of the edge of the jet. For observers inside the jet, $N_{\rm in}(L)$ depends on the distribution of blind spot sizes because the optical depth is large when the shell is illuminated from behind. In this case, the blind spot hides material within it. 
 It was argued by Eichler (2014) that the blind spots size is likely to be larger than $1/\Gamma$. Therefore,  the brightest bursts may well be dominated by $N_{\rm out}(L)$ and possibly some optically thin bursts. If $N_{\rm in}(L)$  is represented by a power law, then it should be softer, more biased towards lower luminosities, and have a steeper power-law index. If at the other extreme the blind spots have all the same size, then  $N_{\rm in}(L)$ is represented by a delta function that peaks at a much lower luminosity  than the maximum luminosity. As it is difficult to calculate the distribution of the sizes of blind spots, we considered both extremes. 
For completeness, we also considered the possibility that the GRB jet is structured and considered the implied luminosity function found by Pescalli et al. (2015). We find that this model can represent the observed data very well. For this, we compared the median luminosities obtained theoretically and from the observed data using the best-fit parameter values obtained by Banerjee et al. (2021). The results are presented in Table \ref{table}. As we can see, the fit is good even for the present case with median luminosity.
 We find that the best value of $\alpha$ calculated in this paper and the best-fit value of $\beta$ and the luminosity break are in agreement with the results of Banerjee et al. (2021) and with other values found in the literature.

\begin{acknowledgements}
We would like to acknowledge the contribution of Prof. David Eichler towards the development of this work, who sadly passed away during the elaboration of this manuscript. We would like to thank  with Avishai Gal-Yam, Richard Ellis, Elena Pian and Paolo Mazzali for useful discussions. We acknowledge funding from Israel Science Foundation and John and Robert Arnow Chair of Theoretical Astrophysics.
\end{acknowledgements}

\end{document}